\documentclass[12pt]{article}
\usepackage{amssymb,amsfonts}
\def \la{\lambda}

\def \di{\displaystyle}

\begin{document}
\baselineskip 18pt

\title{$R$-matrix for asymmetrical $XXZ$ spin 
chain in magnetic field}
\author{P.~N.~Bibikov\thanks{bibikov@PB7855.spb.edu}}

\maketitle

\vskip5mm

\begin{abstract}
The $R$-matrix for asymmetrical $XXZ$ spin chain in magnetic field 
is presented. It depends
non-additively on two spectral parameters. 
\end{abstract}

The following $R$-matrix:
\begin{equation}
R(\hat\la,\hat\mu)=\left(\begin{array}{cccc}
a(\la_1,\mu_1){\rm e}^{\eta(\la_2-\mu_2)}&&&\\
&{\rm sh}\eta&b(\la_1,\mu_1){\rm e}^{\eta(\la_2+\mu_2)}&\\
&b(\la_1,\mu_1){\rm e}^{-\eta(\la_2+\mu_2)}&{\rm sh}\eta&\\
&&&a(\la_1,\mu_1){\rm e}^{\eta(\mu_2-\la_2)}
\end{array}\right),
\end{equation}
where
$a(x,y)={\rm sh}(x-y+\eta)$ and $b(x,y)={\rm sh}(x-y)$,
satisfies the Yang-Baxter equation \cite{1}:
\begin{equation}
(I_2\otimes R(\hat\la,\hat\mu))(R(\hat\la,\hat\nu)\otimes I_2)
(I_2\otimes R(\hat\mu,\hat\nu))=
(R(\hat\mu,\hat\nu)\otimes I_2)(I_2\otimes R(\hat\la,\hat\nu))
(I_2\otimes R(\hat\la,\hat\mu)),
\end{equation}
and the initial condition:
\begin{equation}
R(\hat\la,\hat\la)={\rm sh}\eta I_4,
\end{equation}
where $I_4$ is a $4\times 4$ identity matrix. Here in (1)-(3) $\hat\la=
(\la_1,\la_2)$ is a vector-parameter as well as $\hat\mu$ and $\hat\nu$.

In the limit $\eta\rightarrow 0$ we have:
\begin{equation}
\frac{1}{{\rm sh}(\la-\mu)}\tilde R=I_4+\eta r(\la,\mu)+o(\eta),
\end{equation}
Here $\tilde R=PR$ and $P$
is the standard permutation operator in the space 
${\mathbb C}^2\otimes{\mathbb C}^2$.
The classical $r$-matrix: 
\begin{equation}
r(\la,\mu)=\left(\begin{array}{cccc}
{\rm cth}(\la_1-\mu_1)+\la_2-\mu_2&&&\\
&-\la_2-\mu_2&\frac{\di 1}{\di{\rm sh}(\la_1-\mu_1)}&\\
&\frac{\di 1}{\di{\rm sh}(\la_1-\mu_1)}&\la_2+\mu_2&\\
&&&{\rm cth}(\la_1-\mu_1)-\la_2+\mu_2
\end{array}\right),
\end{equation} 
satisfies the classical Yang-Baxter equation \cite{1}:
\begin{equation}
[r_{12}(\hat\la,\hat\mu),r_{13}(\hat\la,\hat\nu)+r_{23}(\hat\mu,\hat\nu)]+
[r_{13}(\hat\la,\hat\nu),r_{23}(\hat\mu,\hat\nu)]=0.
\end{equation}

According to general scheme of the Quantum Inverse Scattering Method \cite{1}
the $R$-matrix (1) corresponds to asymmetric
$XXZ$ Heisenberg spin chain in external magnetic field. This model is described
by local Hamiltonian ${\cal H}=\sum_{i=1}^N H_{i,i+1}$ with the following
local Hamiltonian density:
\begin{equation}
H_{i,i+1}={\rm e}^{\psi}
\hat\sigma^+_i\hat\sigma^-_{i+1}+{\rm e}^{-\psi}
\hat\sigma^-_i\hat\sigma^+_{i+1}+
\frac{{\rm ch}\eta}{2}\hat\sigma^3_i\hat\sigma^3_{i+1}+
\frac{h}{2}(\hat\sigma_i^3+\hat\sigma_{i+1}^3).
\end{equation} 
Here $N$ is the number of sites of the chain and each $\hat\sigma^k_i$ is the
Pauli matrix $\sigma^k$ acting nontrivially only in the $i$-th space.

In order to connect the $R$-matrix (1) with the local
Hamiltonian density (7) we shall put
$\hat\la=(\la,\frac{\di\la h}{\di\eta{\rm sh}\eta})$,
$\hat\nu=(\nu,\frac{\di\nu h}{\di\eta{\rm sh}\eta})$ and consider elements
of the $R$-matrix as functions of two scalar parameters.
Then the correspondence between $R$ and $H$ is given by the formula:
\begin{equation}
H={\rm sh}\eta L^{-1}(\la,\nu)
\frac{\partial}{\partial\lambda}L(\la,\nu)|_{\la=\nu}-
\frac{{\rm ch}\eta}{2}I_4,
\end{equation}
where $L(\la,\nu)=\tilde R(\la,\nu)$.
The parameter $\psi$ may be expressed from $\nu$ as follows:
$\psi=2\nu\eta$.
According to the Eq. (2) the matrix $L(\la,\nu)$ 
satisfies the standard relation \cite{1}:
\begin{equation}
R(\la,\mu)L(\la,\nu)\otimes L(\mu,\nu)=
L(\mu,\nu)\otimes L(\la,\nu)R(\la,\mu).
\end{equation}

Analysis of the corresponding Bethe ansatz leads to the same formulas as in 
\cite{2}. 
However using the $R$-matrix (1) inside the relation (9) we are directly 
taking into account the interaction with magnetic field.

\end{document}